\documentstyle[12pt]{article}
 \advance\oddsidemargin by -1in
 \advance\evensidemargin
 by -1in
 \marginparsep
 0.4cm
 \advance\topmargin by
 -0.0in
 \makeindex

 \newcommand\noi{\noindent}
 \newcommand\beq{\begin{equation}}
 \newcommand\eeq{\end{equation}}
 \newcommand\beqn{\begin{eqnarray}}
 \newcommand\eeqn{\end{eqnarray}}
 \newcommand{\doublespace} {
\newcommand{\la}{\langle}
 \newcommand{\ra}{\rangle}
 \renewcommand{\baselinestretch} {1.6}
 \large\normalsize}

\begin{document}
\vspace*{1cm}
\hspace*{9cm}{\Large MPI H-V26-1996}
\vspace*{3cm}

 \centerline{\Large \bf Inelastic $J/\Psi$
 Photoproduction off Nuclei:}
\medskip
 \centerline{\Large \bf Gluon Enhancement or 
Double Color Exchange?}

\vspace{.5cm}
\begin{center}
 {\large J\"org~H\"ufner}\\
\medskip
 {\sl Institut f\"ur Theoretische Physik der Universit\"at ,\\
 Philosophenweg 19, 69120 Heidelberg, Germany}
 
\vspace{0.6cm}
 
 {\large Boris~Kopeliovich}
 
\vspace{0.3cm}
 
 {\sl Max-Planck Institut f\"ur
 Kernphysik, Postfach
103980, \newline
 69029 Heidelberg, Germany}\\
 
 
 and
 
{\sl Joint Institute
 for Nuclear Research, Dubna, 141980
 Moscow Region, Russia}

\vspace{0.3cm}
 
 {\large Alexei Zamolodchikov}
 
\vspace{0.3cm}
 
 {\sl Laboratoire de Physique
 Math\'ematique,
Universit\'e
 Montpellier
 II\\ Pl.E.Bataillon, 34095 Montpellier,
France}
 
\end{center}

\vspace{.5cm}
\begin{abstract}

 The nuclear enhancement observed in
 inelastic photoproduction of
$J/\Psi$
 should not be interpreted as evidence
 for an increased gluon
density in
 nuclei. The nuclear suppression 
of the production rate due
 to initial and final
state interactions
 is calculated and a novel two-step color
 exchange
process is proposed, which is
 able to explain the data. 
 
\end{abstract}

\newpage
 
\doublespace
\noi
{\large\bf 1. Introduction}
\medskip

 Measuring inelastic photoproduction of
 $J/\Psi$ off heavy nuclei, $\gamma A
\to
 J/\Psi X$, the EMC collaboration
 \cite{emc} and later the NMC
collaboration \cite{nmc} have observed an
 enhancement of the production rate
compared to a proton or a light nuclear
 target:  the experimental ratio of
the
 cross sections per nucleon on tin to
 carbon exceeds the value one by about $10\%$. 
If this enhancement is fully
 attributed to an enhancement of
the
 gluon density
--  as the authors of experiment \cite{nmc} do -- 
the gluon density in heavy nuclei should 
be larger by a similar value of a
few
 percent, which is perfectly compatible
 with the theoretical
expectations.

 This conclusion is dangerous, however,
 since it neglects
initial and final
 state interactions due to the presence
 of the other
nucleons. The data from
 the E772 experiment \cite{e772} for
 $J/\Psi$
production in proton-nucleus
 collisions show a suppression, which may
 be
parameterized by $A^{-\epsilon}$,
 where $\epsilon = 0.08$.  There is no
reason, why this kind of initial/final
 state interaction should not be
present in photoproduction.
 Therefore, the observed enhancement in
 the
$Sn/C$ ratio in $\gamma\to J/\Psi$
 production of the order of $10\%$ has to
be seen in the light of an expected
 $20\%$ suppression due to initial/final
state interaction.  It is theoretically
 unacceptable to claim that the
discrepancy of $30\%$ is attributed to a
 modified gluon distribution.
Other
 mechanisms have to be looked for.  This
 is the aim of the present
paper, which is
 organized as follows.

 In the next section the
triple-Regge
 approach to inelastic photoproduction of
 charmonium is
developed.  Since the
 charmonium is produced diffractively via
Pomeron exchange, we conclude that
at $x_1\to 1$ the reaction is sensitive to the
double-gluon distribution function (similarly
as in elastic
photoproduction),
 rather than to the single gluon density
 \cite{csm}. Then, in
section 3, 
the suppression factor for inelastic
photoproduction of $J/\Psi$
 off nuclei is calculated using Glauber's eikonal
approximation for the final state
 interaction.  The formulas are
derived for the first time.  
Since Glauber's approximation assumes lack of formation
time we consider the color-singlet model for
$J/\Psi$ photoproduction in section 4. 
We demonstrate that contrary to the wide spread opinion
the time interval for evaporation of color is quite
short, about $1\ fm$ in the energy range of the NMC experiment.
Therefore Glauber's model can be used in this case as well.
Inelastic shadowing corrections to Glauber's approximation 
are considered in section 5. We demonstrate that this 
corrections make the nuclear medium substantially more transparent
for charmonium: the effective absorption cross section turns out
to be only about $60\ \%$ of that for $J/\Psi$.
In section 6 we present the novel mechanism for diffractive
photoproduction of charmonium off nuclei:  the
photon interacts
 with a bound nucleon via color-exchange and
produces a
 $c\bar c$ pair in a color-octet state.
 The pair propagates
through nuclear
 matter and looses energy via 
 hadronization. Then, with another
nucleon the
 $c\bar c$ pair exchanges color
and converts to a color-singlet state.  Such
 a process leads to $J/\Psi$'s
with $x_1
 < 1$.  Since this mechanism relies upon
 two interactions at least,
its cross
 section may have an
 $A^{\alpha}$-dependence with $\alpha >
 1$.
Interplay of this mechanism with
 the standard multiple scattering one
nicely explains the NMC data.
\bigskip 

\noi
{\large\bf 2. Glauber theory for inelastic
$J/\Psi$ photoproduction off nuclei}
\medskip
 
 The inelastic photoproduction cross
section of a $J/\Psi$ on a nucleus can
 be represented in the form
 
\beq
 \frac{d\sigma^{\gamma A}_{\Psi}}{dx_1}
 =
S^{\gamma
 A}_{\Psi}\ A\
 \frac{d\sigma^{\gamma N}_{\Psi}}{dx_1}\ .
\label{4}
\eeq
 
 \noi
In this section we derive the nuclear suppression factor
 $S^{\gamma A}_{\Psi}$ 
within
 Glauber theory. There are two types of
amplitudes
 which
 contribute to the inelastic
 photoproduction of a
$J/\Psi$ off
 nuclei:
 
 I) The incoming photon does not interact in the
 nucleus
until the point
 with
 coordinates $(b,z)$ (the impact
 parameter and the
longitudinal
 coordinate, respectively), where it interacts
 inelastically, $\gamma N \to
J/\Psi X$,
 and produces the $J/\Psi$ with the
 detected final momentum,
which leaves
 the nucleus without inelastic
 interactions;
 
 II) prior the
inelastic
 interaction at
 the point $(b,z)$ the photon produces
elastically a $J/\Psi$
 at the point
 $(b,z_1)$, $\gamma N \to J/\Psi N$, 
with $z_1 < z$. The $J/\Psi$
propagates 
 through
 the nucleus without interaction up to the point
$(b,z)$, 
 where it produces the final
 $J/\Psi$ in the
 reaction $J/\Psi N \to
J/\Psi X$.
 
 In both cases the
 longitudinal momentum
 transfer at the
point $(b,z)$ is so
 large, $q_{in}
 \approx (1-x_1)\nu$, that
 only those
amplitudes interfere, which
 correspond
 to the inelastic production
 of the
$J/\Psi$ with the desired
 momentum on
 the same nucleon.
 
 On the other
hand, the longitudinal
 momentum transfer
 in elastic
 photoproduction at
the point $(b,z_1)$
 may be small at high
 energies,
 
 \beq
 q_{el} =
\frac{Q^2+M_{\Psi}^2}{2\nu}\ .
\label{5}
\eeq
 
 \noi
 Therefore the amplitudes corresponding
 to different
coordinates $z_1$ may
 interfere.
 
 According to our classification of
the
 amplitudes I and II the suppression
 factor $S^{\gamma A}_{\Psi}$ has
three
 terms, $S^{\gamma A}_{\Psi} = (S^{\gamma
 A}_{\Psi})_1 + (S^{\gamma
A}_{\Psi})_{12} +
 (S^{\gamma A}_{\Psi})_2$, where the
 first and the third
terms correspond to
 squares of the amplitudes I and II,
 respectively, and
the second one is
 the interference term.
 
 The amplitude I squared
provides
 the
 following suppression factor
 
 \beq
 \left(S^{\gamma
A}_{\Psi}\right)_1 = {1\over A}\ 
 \int d^2b \int\limits_{-\infty}^{\infty}
 dz\
\rho(b,z)\ \exp\left[-\sigma^{\Psi N}_{in}\
 T_z(b)\right]\ .
\label{6}
\eeq
 
 \noi
 Here
 $T_z(b)=\int_z^{\infty}dz'\rho(b,z')$,
 where
$\rho(b,z)$ is the nuclear density
 normalized to $A$.  The nuclear
thickness function is $T(b) = T_{z\to
-\infty}(b)$.
 
 The interference
term
 has a form
 
 \beqn
 \left(S^{\gamma A}_{\Psi}\right)_{12} &=&
 -
2\
 \left(\frac{1}{2}\sigma^{\Psi N}_{tot}\right)\
{1\over A}\ \int d^2b\
\int\limits_{-\infty}^{\infty} dz
 \rho(b,z)\ \exp\left[- {1\over 2}
\left(\sigma^{\Psi N}_{in} - \sigma^{\Psi N}_{el}\right)\
 T_z(b)\right]
\nonumber\\
 &\times &
 \int\limits_{-\infty}^{z} dz_1\ \rho(b,z_1)\
\cos[q_{el}(z-z_1)]\
 \exp\left[-{1\over 2}\sigma^{\Psi N}_{tot}\
T_{z_1}(b)\right]
\label{7}\ .
\eeqn
 
 This term has a negative sign because
 the amplitude II plays the
role of an
 absorptive correction to amplitude I.
 
 To calculate the
contribution of the
 amplitude II squared we can use the
 results of ref.
\cite{hkn1,hkn2} for
 quasielastic photoproduction of vector
 mesons on
nuclei,
 which we should apply
 to production of the $J/\Psi$ prior the
point
 $(b,z)$. However, as different
 from \cite{hkn1,hkn2}, we should
include both
 coherent (small $q_T$) and incoherent
 (large $q_T$) interactions
of the
 $J/\Psi$ in the nucleus.  The result
 reads,
 
 \beqn
\left(S^{\gamma A}_{\Psi}\right)_{2} &=&
 {1\over A}\int d^2b
\int\limits_{-\infty}^{\infty}
 dz\ \rho(b,z)\
 \left\{\sigma_{el}^{\Psi
N}\
 \int\limits_{-\infty}^{z}dz_1\ \rho(b,z_1)\
 \exp\left[-
\sigma_{in}^{\Psi N}
 T_{z_1}(b)\right]
 \right.
 \nonumber\\
 & + &
{1\over 2A}\
 \sigma^{\Psi N}_{tot}\
 (\sigma^{\Psi N}_{in} -
 \sigma^{\Psi
N}_{el})\
 \int\limits_{-\infty}^{z} dz_1\
 \rho(b,z_1)\
 \exp\left[-
{1\over 2}\sigma^{\Psi N}_{tot}\
 T_{z_1}(b)\right]
 \nonumber\\
 &
\times&
 \left.
 \int\limits_{z_1}^{z} dz_2\
 \rho(b,z_2)\
\cos\left[q_{el}(z_2-z_1)\right]\
 \exp\left[- {1\over 2}
\left(\sigma^{\Psi
 N}_{in}\ -
 \sigma^{\Psi N}_{el}\right)\
 T_{z_2}(b)
\right]\ \right\}\ .
\label{8}
\eeqn
 
 At low energies, when $q_{el} \gg 1/R_A$
 only $\left(S^{\gamma
A}_{\Psi}\right)_1$ and the first term
 in curly brackets in (\ref{8})
survive
 and lead to the nuclear suppression factor
 
 \beq
 S^{\gamma
A}_{\Psi}|_{q_{el}\gg 1/R_A} =
 \frac{1}{\sigma^{\Psi N}_{in}}\
{1\over A}\ \int d^2b\
\left\{\frac{\sigma^{\Psi N}_{tot}}
 {\sigma^{\Psi N}_{in}}
\left[1-e^{-\sigma_{in}^{\Psi N}T(b)}\right]
 - \sigma^{\Psi N}_{el}\ T(b)\
e^{-\sigma_{in}^{\Psi N}T(b)}\right\}\ .
\label{9}
\eeq
 
 At high energies, in the limit $q_{el}
 \to 0$ the sum of
expressions
 (\ref{7})-(\ref{9}) simplifies very
 much,
 
 \beq
S^{\gamma
 A}_{\Psi}|_{q_{el}\ll 1/R_A} =
{1\over A} \int d^2b\
T(b)e^{-\sigma_{in}^{\Psi N}T(b)}\ .
\label{10}
\eeq
 
 Using the smallness of $\sigma^{\Psi
 N}_{tot} \approx 5\ mb$ we
can
 expand
 (\ref{9}) as $S^{\gamma
 A}_{\Psi}|_{q_{el}\ll 1/R_A} \approx 1
-
 {1\over 2} \sigma^{\Psi N}_{tot}\
 \langle T\rangle$, where the mean
nuclear
 thickness $\langle T\rangle =
 A^{-1} \int d^2b\ T^2(b)$. In the approximation eq. 
(\ref{10}) we have kept
only the first order of the expansion
 parameter $\sigma^{\Psi N}_{tot}
\langle
 T\rangle$, and neglected
 $\sigma^{\Psi N}_{el}\langle
 T\rangle$, which is of the
second
 order. To the same order the
 low-energy limit eq.~(\ref{9}) leads to
 $S^{\gamma A}_{\Psi}|_{q_{el}\gg 1/R_A}
 \approx 1 - \sigma^{\Psi
N}_{tot}\
 \langle T\rangle$.

 Thus, the nuclear suppression factor at
high energy
 turns out to be smaller than
 that at low energy.  This result has a
natural
 space-time interpretation: at high
 energies a photon may interact strongly
via its hadronic
 fluctuations. The lifetime of
the
 $c\bar
 c$ fluctuation of the photon,
 called coherence time, is
$t_c\sim
 1/q_{el}$.
 A fluctuation created
 long in advance 
 propagates
 and attenuates through the
 whole nuclear
thickness, as is
 explicitly
 exposed in (\ref{10}).  On
 the other hand, at
low energy the photon
 converts
 to a $c\bar c$ inside the nucleus
 just prior to
the $J/\Psi$ production.  So the
 $c\bar
 c$ pair propagates only through
 about a
half of the nuclear thickness.
 Such
 a contribution corresponds to
$(S^{\gamma A}_{\Psi})_1$ given by
 (\ref{6}).
 There is, however, a
possibility to produce $J/\Psi$
 ''elastically'' in
 $\gamma N \to J/\Psi
N$ prior the inelastic rescattering
 $J/\Psi N \to
 J/\Psi X$.  This
correction is suppressed by factor
 $\sigma^{\Psi N}_{el}/
 \sigma^{\Psi
N}_{in}$ and is also included in eq.~(\ref{9}).
 
 Expressions
 (\ref{6}) -
(\ref{8}) give the correct 
 extrapolation between the low-energy
eq.~(\ref{9}) and
 the
high-energy eq.~(\ref{10})
 limits and are quite complicated.  If one uses 
 the
 same
approximation
 $\sigma^{\Psi N}_{tot}\ \langle T\rangle
 \ll
1$ and expands eqs.~(\ref{6}) - (\ref{8})
in this small parameter up to
first order (compare with quasielastic
 photoproduction of $J/\Psi$ off
nuclei
 \cite{kz,benhar}), one finds
 
 \beq
 S^{\gamma A}_{\Psi} \approx
 1 -
{1\over 2}\sigma^{\Psi N}_{tot}\
 \langle T\rangle\
 \left[1 +
F_A^2(q_{el})\right]\ ,
\label{11}
\eeq
 
 \noi
where the effective longitudinal nuclear formfactor
 squared
is defined as
 
 \beq
 F_A^2(q_{el}) =
 \frac{1}{A\ \langle T\rangle}\
\int d^2b\
 \left|\int\limits_{-\infty}^{\infty}
 dz\ \rho(b,z)\
\exp(iq_{el}z)\right|^2\ .
\label{12}
\eeq
 
 \noi
One can approximate
 $F_A^2(q_{el})$ quite well using
a Gaussian parameterization
 instead of the more realistic Woods-Saxon
 form for the
nuclear
 density \cite{jager}
and obtains 
$F_A^2(q_{el}) = \exp(- R_A^2\ q_{el}^2)/3$, 
where $R_A^2$ is
 the mean squared charge radius
of the nucleus \cite{jager}.
  Note that
 $S^{\gamma A}_{\Psi}$ is
independent of  $x_1$, therefore this mechanism of
nuclear
 suppression keeps the same
 $x_1$-distribution of the cross
section
 as on a
 free nucleon and cannot account for the data
(cf. section 4).
 
\medskip
 
 There is another possibility within the
 standard Glauber
approach to produce the
 $J/\Psi$ in two-step process on a
 nucleus: the photon produces
 the $J/\Psi$ inelastically on one
 nucleon
with a subsequent diffractive
 scattering of the charmonium on another
nucleon.  Inelasticities 
 of these two interactions should
be
 adjusted to provide the desired $x_1$ 
for the final $J/\Psi$.
 The contribution of this
mechanism to
 the cross section is evaluated in the
 Appendix and a value of less than $2\%$
is found, which 
we neglect 
 compared to the main mechanism
 considered above.
\bigskip

\noi
{\large\bf 3. Triple-Regge formalism for
$J/\Psi$ photoproduction.}
\medskip

As soon as the nuclear suppression factor 
$S^{\gamma A}_{\Psi}$ is calculated
within Glauber approximation, one can predict the
nuclear cross section provided that the $J/\Psi$ photoproduction
cross section on a free nucleon is known. We review briefly
in this and the next sections the well accepted mechanisms
in order to justify the usage of Glauber 
model and to get a reasonable 
parameterization for the cross section.

 In the standard triple-Regge
 formalism
 (see for instance \cite{3r}) the
 diffractive inelastic
 photoproduction of
$J/\Psi$ is described by the
 triple-Pomeron graph,
 depicted in
 Fig.~1.

\begin{figure}[tbh]
\includegraphics{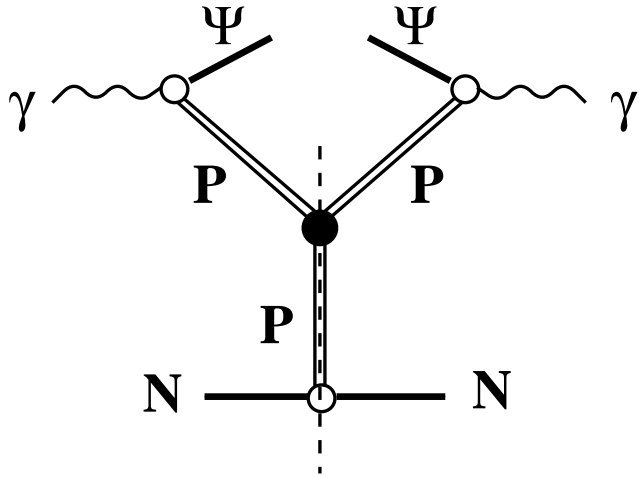}
\begin{center}
\vspace{6cm}
\parbox{13cm}
 {\caption[Delta]
 {The triple-Pomeron graph for the cross
 section of the reaction $\gamma N \to
 J/\Psi X$.  The dashed line shows
 that
 only the absorptive part of the
 amplitude is included.}
\label{fig1}}
\end{center}
\end{figure}

 The corresponding differential cross
 section reads

 \beq
 \frac{d\sigma(\gamma N \to J/\Psi X)}
 {dx_1\ dt} =
 G(\gamma N \to J/\Psi
 X)
 \frac{\exp(- B^{\Psi p}_{in}p_T^2)}
 {(1-x_1)^{2\alpha_P(t)-1}}\ .
\label{1}
\eeq

 \noi
 Here $x_1$ and $t$ are the relative share of the
photon light-cone momentum carried by the $J/\Psi$ 
and the 4-momentum squared of
 the produced $J/\Psi$, respectively.
 $\alpha_P(t)$ is
the Pomeron
 trajectory and $B^{\Psi p}_{in}$ is the
 slope parameter for
inelastic $J/\Psi-p$
 collisions. The factor
 $G(\gamma N \to J/\Psi X)$
includes the
 triple-Pomeron
 coupling as well as the
 $P\gamma\Psi$ and the
$PNN$ vertices.
 It can be
 evaluated using the data from
 hadronic
interactions, assuming
 factorization of
 the Pomeron.

 \beq
 G(\gamma
N \to J/\Psi X) =
 G(pp \to p X)\ \frac
 {\sigma(\gamma p \to J/\Psi
 p)\
B_{el}^{\Psi p}}
 {\sigma_{el}^{pp}\ B_{el}^{pp}}\ .
\label{2}
\eeq

 The momentum transfer squared $t$ is
 rather large in $\gamma N \to
J/\Psi X$,
 $|t|_{min} = (M^2_{\Psi} +
 x_1Q^2)(1-x_1)/x_1$, where
$x_1=p_{\Psi}^+/p_{\gamma}^+$ ($ \approx
 x_F$ at $x_1 \to 1$) is the fraction
of
 the photon light-cone momentum carried
 by the $J/\Psi$.  $|t_{min}|$ may
reach
 a few $GeV^2$ dependent on $x_1$.  The
 Pomeron trajectory
$\alpha_P(t)$ is
 known
 to approach the value of 1 at high $|t|$
\cite{bfkl}, therefore we take
 $\alpha_P(t) = 1$.

 Combining all the
factors we get an
 estimate for the $x_1$-distribution of
 the inclusive
photoproduction of
 $J/\Psi$ on a proton, integrated over
 transverse
momenta

 \beq
 \frac{d\sigma(\gamma N \to J/\Psi X)}
 {dx_1} =
\frac{B^{\Psi p}_{el}}
 {B^{pp}_{el}\ B^{\Psi p}_{in}}
 \
\frac{\sigma(\gamma p \to J/\Psi p)}
 {\sigma^{pp}_{el}}\
 \frac{G(pp \to
pX)}
 {1-x_1}\ .
\label{3}
\eeq

 The effective triple-Pomeron constant
 $G(pp \to p X) = 3.2\ mb/GeV$
is fixed
 by an analysis of data on $pp \to
 pX$ \cite{3r}.  The elastic $pp$
cross section and
 the slope parameter are taken as
 $\sigma_{el}^{pp}=8\ mb$
and $B^{pp}_{el}
 \approx 10\ GeV^{-2}$. The slope of
the $p_T$-distribution of
elastic $J/\Psi$
 photoproduction is assumed $B^{\Psi
 N}_{el}\approx
B^{pp}_{el}/2$, since the
 slopes are related to the nucleon and
 the $J/\Psi$
form factors respectively.
 On the other hand, in the inelastic
photoproduction of $J/\Psi$ the nucleon
 is destroyed and its formfactor does
not
 contribute to the effective slope (the
 same as in the case of $pp\to
pX$
 reaction, where the slope $B^{pp}_{in}\approx 
4\ GeV^{-2}$ \cite{3r}). Therefore $B^{\Psi N}_{in}$
is only due
 to the charmonium formfactor.  It was
 measured in the EMC
experiment
 \cite{emc-fe} to be $B^{\Psi p}_{in} = 0.58
 \pm 0.07\ GeV^{-2}$,
but the NMC
 experiment \cite{jong} led to $B^{\Psi
 p}_{in} = 0.23 \pm 0.02\
GeV^{-2}$.
For a mean photon energy $\langle \nu
 \rangle \approx 100\
GeV$, the elastic
 photoproduction cross section is
 $\sigma(\gamma p \to
J/\Psi p) \approx
 12\ nb$ according to the EMC
 measurements \cite{emc-fe},
but is
 $\approx
 27\ nb$ according to the NMC
results
 \cite{chiara}. In view of the 
 uncertainties in the measured values of
the
 elastic photoproduction cross section
 and the inelastic slope we fix
them at a
 middle value $\sigma(\gamma p \to J/\Psi
 p) = 20\ nb$ and
$B^{\Psi p}_{in}
 = 0.45\ GeV^{-2}$.

 The the cross
section formula
 (\ref{3}) is calculated with the 
 parameters given above and is compared 
with the data from the NMC
 experiment \cite{chiara} in Fig.~2.
A good agreement is found 
both in shape and normalization.  

\begin{figure}[tbh]
\includegraphics{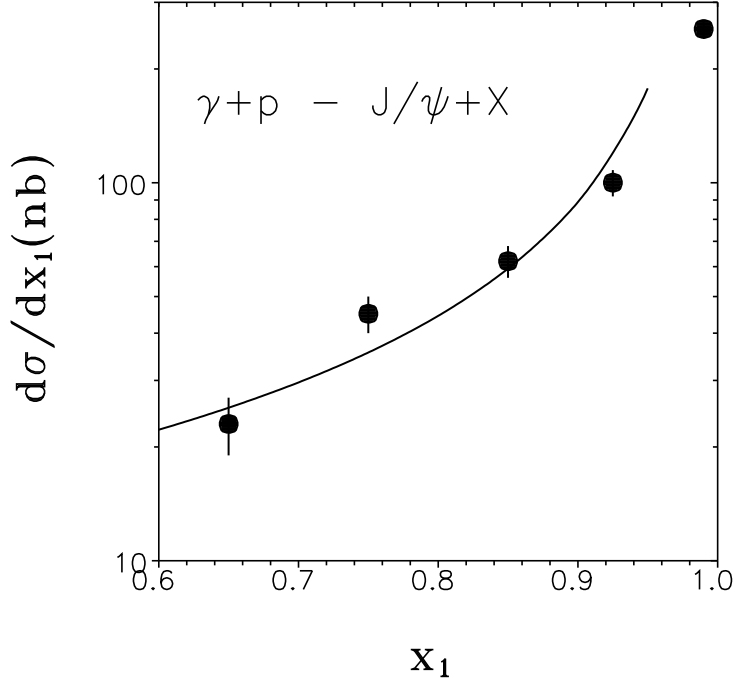}
\begin{center}
\vspace{9cm}
\parbox{13cm}
 {\caption[Delta] {The data from the NMC
 experiment
 \cite{nmc}
 for inelastic
 $J/\Psi$ photoproduction on protons.
 The solid
 curve shows
 predictions based
 on the triple-Pomeron phenomenology.}
\label{fig2}}
\end{center}
\end{figure}

Note that the triple-Pomeron
is known to dominate in the inclusive cross section
of $pp$ interaction only at $x_1 > 0.9$ \cite{3r}.
The nondiffractive RRP triple-Reggeon graphs
are important at smaller $x_1$.  
OZI rule suppresses the RRP term
for $J/\Psi$ interactions, therefore 
the triple Pomeron contribution may be important
down to lower $x_1$. Nevertheless, the triple-Pomeron
 phenomenology does not necessarily
describe the data for $x_1 < 0.8$.
 Particularly, the longitudinal
formfactor of the $\gamma \to J/\Psi$
 vertex may substantially diminish the
cross section at mid $x_1$.
Yet we need a parameterization
 of the
data for further applications,
 and Fig.~2 shows that eq. (\ref{3})
reproduces the data well down to
 $x_1=0.6$.

 Usually the
inelastic photoproduction of
 $J/\Psi$ is treated perturbatively, 
within the so called
 color-singlet model \cite{csm}.  In
that approach the cross section is
predicted to be proportional to the low-$x$
gluon density distribution in the
 proton.  Although the Pomeron by definition
includes all possible kinds of
 perturbative graphs, but the
ones of the color-singlet
model are not the
 leading contributions at $x_1 \to 1$.
Therefore the spectrum of $J/\Psi$
 predicted by the color singlet model
 does not peak at
$x_1 \to 1$, while the
 triple-Pomeron contribution does.  The
 dominant
mechanism comes from the
 ladder-type graphs with excitation of
 the proton.
Correspondingly, the cross
 section is proportional to the
 double-gluon
density, which is
 approximately equal to the single-gluon
 density
squared.

 We conclude that for $x_1 \to
 1$ the inelastic
 photoproduction of $J/\Psi$ 
cannot be used as a probe for the
 single-gluon distribution,
but it might
 be possible at mid $x_1$-values.
\bigskip

\noi
{\large\bf 4. Color-singlet model}
\medskip

A popular approach to inelastic diffractive photoproduction
of $J/\Psi$ on a nucleon target is the 
color-singlet model (CSM) \cite{csm}. It is somewhat similar to
the mechanism suggested in section 6, except that
the color-octet $c\bar c$ pair becomes colorless via
radiation of a gluon, rather than by means of a color-exchange 
rescattering. 

Radiation of a gluon takes some time, which may be important
if the process takes place in a nuclear environment.
Indeed, rescatterings of the color-octet $c\bar c$ pair
may induce additional energy loss, and may result in an effective
A-dependence. Therefore we need to understand the space-time
pattern of gluon radiation in the CSM.

Sometimes the production of a charmonium in CSM is
associated with a two-step space-time development: 
first the color-octet $c\bar c$ pair is produced, 
then it propagates and finally 
converts into a color-singlet state via radiation
of a gluon. The second stage is usually assumed to take
a long time.

We do not think that such a treatment is possible in view of a
principal quantum-mechanical uncertainty: analogous to the
standard electromagnetic bremsstrahlung one cannot
say whether the gluon/photon has been radiated before or after
the interaction. Both Feynman graphs contribute to
the cross section (actually both are taken into account
in perturbative calculations \cite{csm}). 
 One can understand this uncertainty as related to 
the lifetime of a fluctuation on the initial particle
containing the radiated gluon and the color-octet
$c\bar c$ pair. This lifetime given by the 
energy-denominator equals to

\beq
t_r = \frac{2\nu}{M^2}\ ,
\label{12a}
\eeq

\noi
where $M$ is the effective mass of the fluctuation,

\beq
M^2 = \frac{M^2_{c\bar c}}{x_1} +
\frac{k_T^2}{x_1(1-x_1)}\ .
\label{12b}
\eeq

\noi
Taking the mass of the $c\bar c$ pair $M_{c\bar c} \approx
M_{\Psi}$ we get the following expression for the radiation
time

\beq
t_r = \frac{2\ \nu\ x_1(1-x_1)}
{k_T^2 + M_{\Psi}^2(1-x_1)}\ .
\label{12c}
\eeq

\noi
This radiation time  is only about $1\ fm$
in the kinematical region of the NMC experiment. 
One can safely neglect an influence of nuclear medium on
the process of hadronization for such a short time interval.
Therefore the multiple scattering approach developed in sections
2 can be applied in the case of CSM.

Note that even at very high energies, when $t_r$ becomes long,
the effects of induced radiation by the color-octet $c\bar c$ pair 
during radiation time
can be neglected since $2\kappa_{ind}/M_{\Psi}^2 \ll 1$,
where $\kappa_{ind} \leq 1\ GeV^2$ is the density of energy loss
via induced radiation.
 \bigskip

\noi
{\large\bf 5. Beyond Glauber approximation}
\medskip

Considering photoproduction of $J/\Psi$ in the quark
representation one understands that a colorless 
$c\bar c$ wave packet, rather than a $J/\Psi$, 
is produced and propagates 
through the nucleus, but the $J/\Psi$ may be
formed far away from the nucleus if the energy is high.
Due to quark-hadron duality one can consider the same effects
taking into account photoproduction of higher diffractive
excitations of the charmonium, $\Psi'$, $\Psi''$ etc., as well as
all diagonal and off-diagonal diffractive rescatterings
of these states in nuclear matter. 
The corresponding corrections are know 
as inelastic shadowing \cite{gribov}.
The simplest, but quite
accurate two-coupled-channel approximation including $J/\Psi$ and
$\Psi'$ was considered in \cite{hk}. It was found that the
the $c\bar c$ wave packet produced in $pp$ interaction is
such a combination of $J/\Psi$ and $\Psi'$, which is quite close
to the eigenstate of interaction having the minimal possible 
interaction cross section.
This fact explains particularly why $J/\Psi$ and $\Psi$ experience
similar nuclear attenuation (cf. the E772 experiment \cite{e772}).

We assume that the ratio of the photoproduction amplitudes

\beq
R_{\gamma p} =
\frac{\langle\Psi'|\widehat f|\gamma\rangle}
{\langle\Psi|\widehat f|\gamma\rangle}\ ,
\label{11a}
\eeq

\noi
where $\widehat f$ is the operator of diffractive amplitude, 
is close to one. In $pp$ scattering the corresponding value of 
$R_{pp}$ is measured to be $0.5$ (see in \cite{hk}). 
This assumption on $R_{\gamma p}$ can be 
motivated by the similarity of the $c\bar c$ components of the 
wave functions of a photon and of a gluon.

The solution of evolution equation for
the $c\bar c$ wave packet propagated through nuclear matter,
which incorporates with both effects of the coherence and formation
lengths will be presented elsewhere \cite{hkfuture}, but some
numerical results are published in \cite{moriond}. Here we use 
the approximation of small 
$\sigma^{\Psi N}_{tot}\langle T\rangle \ll 1$. Then 
one can modify the Glauber model expression (\ref{11}) as 
follows \cite{moriond}

 \beq
S^{\gamma A}_{\Psi} \approx
1 - \sigma_{eff}T_{eff}
\approx exp(- \sigma_{eff}T_{eff})\ ,
\label{11b}
\eeq

\noi
where

\beq
T_{eff} = {1\over 2} 
 \langle T\rangle
 [1 + F_A^2(q_c)]
\label{11c}
\eeq

\noi
is according to eq.~(\ref{11}) the effective
nuclear thickness covered by the charmonium, which
is controlled by the coherence length $l_c
\approx 1/q_{el}$ \cite{hkn2}, and 

\beq
\sigma_{eff} = \sigma^{\Psi}_{in}
[1 + \epsilon\ R\ 
F_A^2(q_f)]
\label{11d}
\eeq

\noi
is the effective absorptive cross section for charmonium,
which is controlled by the formation length for the charmonium
wave function, $l_f \approx 1/q_f$, where

\beq
q_f =
\frac{M^2_{\Psi'}-M^2_{\Psi}}{2x_1\nu},\
\label{11e}
\eeq

\noi
is the longitudinal momentum transfer in the
off-diagonal reaction $\Psi N \rightleftharpoons
\Psi' N$. 

The new parameter $\epsilon$ in eq.~(\ref{11d})
is

 \beq
 \epsilon = \frac{\la\Psi'|\widehat f|\Psi\rangle}
{\la\Psi|\widehat f|\Psi\rangle}\ ,
\label{11f}
\eeq

\noi
and is estimated in \cite{hk} in 
the oscillator approximation,
$\epsilon = -\sqrt{2/3}$.
Remarkably, with these parameters 
at high energies $\sigma_{eff} \approx
0.6\ \sigma^{\Psi N}_{in}$, i.e.
the $J/\Psi$
attenuates in nuclear matter substantially
less than  predicts Glauber model.
This may be understood as a natural
consequence of color transparency.

We calculate ratio of production rates of
$J/\Psi$ for tin to carbon using eq.~(\ref{11b}).
The results for $\nu = 70$ and $100\ GeV$ are plotted 
in Fig.~3 in comparison with the data from the
NMC experiment \cite{nmc}. We see that the discrepancy
with the inelastic photoproduction data ($x_1 < 0.85$)
is so large that it is difficult to explain it by
a few percent enhancement of the gluon density in
heavy nuclei. One has to look for another explanation
of the EMC-NMC effect.

\begin{figure}[tbh]
\includegraphics{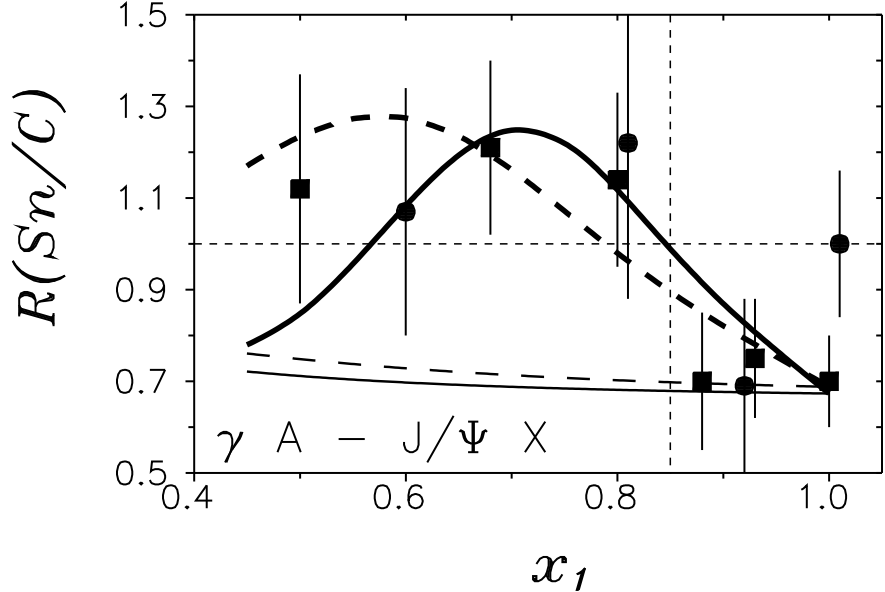}
\begin{center}
\vspace{8cm}
\parbox{13cm}
 {\caption[Delta]
 {The data from the NMC experiment
 \cite{nmc} for the ratio of inelastic $J/\Psi$
 photoproduction rates on tin to carbon, measured 
with $200\ GeV$
 (squares) and $280\ GeV$ (circles)
 muons. The thin solid and dashed curves
 correspond to the Glauber-Gribov mechanism
for nuclear suppression of $J/\Psi$ and $\nu= 100$
and $70\ GeV$, respectively.
The thick solid and dashed curves
 show predictions based on the
 proposed novel mechanism of diffractive
 interaction with a nucleus at $\nu=100$
and $70\ GeV$, respectively. For the origin 
of the dividing line at $x_1=0.85$ see text.}
\label{fig3}}
\end{center}
\end{figure}


\noi
{\large\bf 6. Novel mechanism of $J/\Psi$
photoproduction in nuclear matter}
\medskip
 
 According to the arguments in section 3 the
photoproduction of $J/\Psi$ on a nucleon proceeds via
Pomeron exchange, which  in the language of perturbative QCD
can be
interpreted as a colorless gluonic exchange,
containing at least two gluons in
the $t$-channel.  In a nuclear interaction 
this colorless exchange  can be
split into two
 color-exchange interactions with
 different nucleons, keeping the
 beam and the nuclear remnants in
the final state colorless
\cite{kl,lnpi}. 
 
 The same physics can be represented as a
 consequence of
unitarity: the
 imaginary part of the elastic forward
 scattering
amplitude of $J/\Psi$ on a
 nucleon is a product of an inelastic
 $J/\Psi\ N$
amplitude with itself
 time-conjugated (up to small elastic and
 diffractive
corrections).
 In the case of a nuclear target one has to
consider also 
products of two inelastic
 amplitudes on different nucleons.
 
Thus, the
 $J/\Psi$ photoproduction on a
 nucleus can result from the
photoproduction
 on a bound nucleon of a
 $c\bar c$ pair in a color-octet
state,
 which
 propagates through 
 nuclear matter loosing energy
via
 hadronization
until another
 color-exchange interaction 
inside the nucleus
converts the pair back to
 a colorless state.
Actually, this is
 a specific QCD mechanism of diffractive
 excitation of the
nucleus.  The energy
 of
 the $J/\Psi$ (and the excitation
 energy of the
nucleus) is less than the
 photon energy by the amount lost via
hadronization by $c\bar c$ pair while
 it
 is in the color octet state. We
assume
 a constant energy loss per unit
 of
 length $dE/dz = -\kappa$, what
is true
 both for the color string
 \cite{cnn} and
 for the bremsstrahlung
\cite{nieder}
 mechanisms of
 hadronization. In this
 case we have
 
\beq
 x_1 = 1 - \frac{\kappa}{\nu}\Delta z
\label{13}\ ,
\eeq
 
 \noi
 where $\Delta z$ is the longitudinal
 distance covered by
the
 $c\bar c$ pair
 in the color-octet state.
 
 The correction from this
mechanism
 to the cross section for
nuclear photoproduction can be
calculated as follows

 \beqn
 \Delta\left(\frac{d\sigma^{\gamma
 A}_{\Psi}}{dx_1}\right) &=&
2\pi\
 \frac{\nu}{\kappa}\
 B^{\Psi N}_{el}\
 \sigma(\gamma N \to J/\Psi N)\
\int
 d^2b\
 \int\limits_{-\infty}^{\infty}
 dz_1\ \rho_A(b,z_1)\
\int\limits_{z_1}^{\infty}dz_2\
 \rho_A(b,z_2)\nonumber\\ & &
\delta\left[z_2 - z_1 -
 (1-x_1)\frac{\nu}{\kappa}\right]\
\exp\left\{-\sigma^{\Psi N}_{in}\left[T(b)-
 T_{z_1}(b) + T_{z_2}(b)
\right]\right\}\ .
\label{14}
\eeqn
 
 \noi
 The derivation of this expression
 includes following
ingredients.  We
 assume that the photon energy is
 sufficiently high, $\nu
\gg
 M_{\Psi}^2R_A/2$ to allow the photon to
 convert into the $c\bar c$
pair
 long in
 advance. At the first
 interaction point
$(b,z_1)$ the $c\bar c$ pair converts into the
 color octet state,
corresponding 
 to the
 cut Pomeron with cross section
 $\sigma_{in}^{\Psi N}$.
It is
 important
 that no
 longitudinal momentum is
transferred to the $c\bar c$ pair
 at
 this point. All the loss of
 longitudinal momentum,
$(1-x_1)\nu$,
 is
 due to the hadronization, until the
 last
 interaction
at point $(b,z_1)$, where
 the color octet $c\bar c$ pair 
converts back to
 the
colorless state. This
 interaction
 proceeds with the same
 cross section as the very
first one,
 except 
for a factor
 $1/8$ (no summation
over
 final colors of the
 $c\bar c$ in this
 case). We also use the relation
$\sigma_{el}^{\Psi N} =
 (\sigma^{\Psi
 N}_{tot})^2/16\pi B_{el}^{\Psi N}$,
where
 $B_{el}^{\Psi N}\approx 5\
 GeV^{-2}$ is the
 slope of $\Psi N$ elastic
scattering (see section 3.
 The
 $\delta$-function in eq.~(\ref{14})
 takes care of
energy-conservation
 including
 the energy loss (\ref{13})
 during the
intermediate state
 hadronization. We
 also assume that the
 $c\bar c$ pair
does not attenuate when it
 propagates through
 the nucleus in a
 color octet
state.  Actually,
 rescatterings of such a pair
 also may
 cause an
attenuation \cite{kop}, but it
 is small provided
 that the photon
energy is sufficiently high
 $(1-x_1)\nu/\kappa \gg R_A$.
 However, the
$c\bar c$ pair does attenuate
 propagating through the nucleus
 in the
colorless state before the first
 interaction (if $q_{el}R_A \ll
 1$) and after the last one.  
We assume in eq.~(\ref{14}) a long
coherence length and zero formation length.
We neglect the energy variation due to
these effects because the estimate eq.~(\ref{14})
is quite rough and the accuracy of data \cite{emc,nmc}
is low as well.
 
 The only unknown parameter we are left with in
eq.~(\ref{14}) is $\kappa$,
the energy
 loss per unit length. In the string model it
 equals the
string tension.
The color-octet string
 tension $\kappa$ may
substantially exceed the
 known value for
 color-triplet
strings $\kappa_3 \approx
 (2\pi\alpha_R')^{-1} \approx 1\ GeV/fm$
\cite{cnn},
 where $\alpha_R' \approx 1\ GeV^{-2}$ is
 the universal slope
of
 Regge
 trajectories. Indeed, the string
 tension of the color-octet
string
 must
 to be related to the slope of the
 Pomeron trajectory,
$\alpha'_P\approx
 0.2\ GeV^{-2}$, which is much smaller
 than $\alpha_R'$.
Thus, $\kappa_8
 \approx (2\pi\alpha_P')^{-1} \approx 
 5\ GeV/fm$
\cite{lnpi}.
 
 One may also treat the energy loss
 perturbatively as a
result
 of gluon
 bremsstrahlung by the color-octet $c\bar
 c$ pair.  It
was
 demonstrated in
 \cite{nieder} that bremsstrahlung
 provides a
constant
 density of energy loss, similar
 to the string model.  The mean
squared color-octet
 charge is $9/4$ times bigger
 than the color-triplet one, so
is the
 energy loss.  Besides, in the case
 under discussion the
color-charge
 appears during
 the first interaction
 with a bound nucleon at
the time $t=t_1$, then
 propagates for the time,
 $\Delta t = t_2-t_1$ and
disappears  when the
 $c\bar c$
 pair becomes colorless.  Such
 a process with switch-on 
and switch-off causes a double
 bremsstrahlung \cite{nieder} (see also
 \cite{kl}): only
 that part of the
frequency spectrum is
 emitted during a time interval
$\Delta t$,
 which has
 radiation time $t_r =
 2\omega/k_T^2 < \Delta t$,
where
 $\omega$ and $k_T$
 are the energy and
 transverse momentum of the
radiated gluon, respectively.
 At the time 
 $t=t_2$ when the $c\bar c$ pair
 converts to a
colorless state and the
 radiation
 process stops. A new
 radiation
process caused by the charge
 stopping starts, and another piece of
 the
gluon spectrum is radiated,
 identical to the one previously emitted.
 Thus, the
intensity of radiation and
 the
 energy loss are twice as large as in a
single-scattering process.
 Summarizing,
 we expect the energy loss for
gluon
 bremsstrahlung in the
 double-scattering
 process with a color-octet
intermediate
 state to be $4.5$
 times larger than that
 for radiation of a
color-triplet charge,
 produced in
 a single scattering process.
 This
estimate is very close to the one we
 found
 above in the string model, so we fix
for further calculations $\kappa = 5\
 GeV/fm$.
 
 The new mechanism
provides an $x_1$-dependence
 of the cross section
 quite different from the
standard one.
 There is no cross section for 
$x_1 < 1-\kappa R_A/\nu$,
 since
the amount of lost energy is restricted
 by the length of the path of the
color-octet
 $c\bar c$ pair inside the nucleus. This contribution
 does not
peak at $x_1 \to 1$, but may have even a
 minimum due to a longer path of the
colorless $c\bar c$ pair
 in nuclear matter.
 
 The contribution of this
mechanism may
cause an $A$-dependence steeper than $A^1$,
since there is more room for  integration over
longitudinal coordinates $z_1$ and $z_2$. 
For this reason the contribution
of the mechanism under discussion, is important
for heavy nuclei at moderately high $x_1$. We compare the contribution of 
Glauber-Gribov mechanism calculated with eq.~(\ref{11b}) and
the novel one given by eq.~(\ref{14}) to
$J/\Psi$ photoproduction on lead at different photon
energies in Fig.~4. We see a strong energy variation of the
latter contribution, which shifts with energy towards $x_1=1$.
At the same time the energy dependence of nuclear effects provided
by a variation of the coherence and formation lengths is so small
that we plotted only one curve for the Glauber-Gribov
mechanism at $\nu=100\ GeV$.

We see that both mechanisms the Glauber one and
the two-step production are of the same order at $x_1
\approx 0.6 - 0.7$. 

\begin{figure}[tbh]
\includegraphics{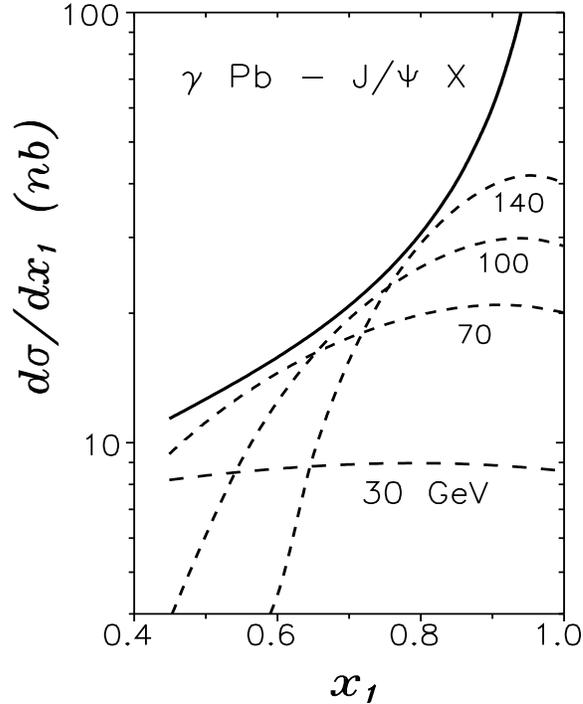}
\begin{center}
\vspace{9cm}
\parbox{13cm}
 {\caption[Delta]
 {The cross section per nucleon for 
$J/\Psi$ photoproduction on
lead. The solid curve
corresponds to Glauber-Gribov mechanism
of nuclear shadowing at $\nu=100\ GeV$.
The novel mechanism contributions at 
$\nu=70$, $100$ and $140\ GeV$ are shown by 
dashed curves.}
\label{fig4}}
\end{center}
\end{figure}
 
The novel mechanism proposed here does not display 
the Feynman scaling typical
for the standard mechanism, but 
 this contribution grows with energy at fixed
$x_1$.
 This is easy to understand: the total contribution
 integrated over
$x_1$ is energy independent, but the
 range of $1-x_1$ shrinks $\propto
1/\nu$.

 With the above value of $\kappa=5\ GeV/fm$ and
 $\sigma_{in}^{\Psi N} = 5\ mb$ we
calculate the contribution of the
 double-color-exchange mechanism
(\ref{14}).  Combining with the 
 Glauber contribution from section 2 we
have
 
 \beq
 \frac{d\sigma^{\gamma A}_{\Psi}}{dx_1} =
 S^{\gamma
N}_{\Psi}\ \frac{d\sigma^{\gamma
 N}_{\Psi}}{dx_1} +
\Delta\left(\frac{d\sigma^{\gamma
 A}_{\Psi}}{dx_1}\right)\ .
\label{15}
\eeq
 
 \noi
The result for the ratio
 of the cross sections for tin to carbon
 is plotted in Fig.~3 for two photon energies $\nu=70$ and
$100\ GeV$ typical for the NMC experiment in
 comparison with
 the data \cite{nmc}.
 
 It should be noted that such a comparison
is legitimate only for $x_1 < 0.85$
 where the experimental 
 resolution of the $J/\Psi$ momentum 
 suffices to
 separate events of inelastic
 photoproduction from the quasielastic ones. 
 For  $x_1 > 0.9$ 
 quasielastic photoproduction of $J/\Psi$
 is supposed
 \cite{emc,nmc} to be
 the dominant
 process.  In this region we should
 compare the data with the
 quasielastic
 photoproduction cross section.
 The corresponding formula is derived in
 \cite{hkn1,hkn2}. It is different from
formulas (\ref{6}) - (\ref{8}) for inelastic
photoproduction, but in the approximation
$\sigma^{\Psi
N}_{tot}\langle T\rangle \ll 1$ looks 
analogous to eq.~(\ref{11b})
 
\beq
 \sigma_{Qel}(\gamma A\to\Psi) \approx
 \sigma(\gamma N \to \Psi N)\
A\
 \left\{ 1 -
 {1\over 2}\sigma^{\Psi N}_{tot}\
 \langle T\rangle\
\left[1 +
 F_A^2(q_{el})\right]\ 
\left[1 + \epsilon\ R\
F_A^2(q_f)\right]
\right\}\ ,
\label{18}
\eeq
 
\noi
The results of calculation with eq.~(\ref{11b}) shown
in Fig.~3 correspond to quasielastic scattering in the limit
$x_1 \to 1$, where the momentum transfer eq.~(\ref{11e})
acquires the value $q_f = (M_{\Psi'}^2-M_{\Psi}^2)/2\nu$.

Fig.~3 shows a good agreement 
with the data both for inelastic and
 quasielastic photoproduction of $J/\Psi$
off nuclei.

It is important to study the energy dependence of the 
effect under discussion, which is expected to 
be interesting. It seems most likely that new data 
on inelastic $J/\Psi$ photoproduction will be taken with the 
HERMES spectrometer at HERA. In Fig.~5 we show predictions
for nuclear suppression/enhancement factor $S^{\gamma A}_{\Psi}$
at $\nu=30\ GeV$ for xenon and nitrogen. 

\begin{figure}[tbh]
\includegraphics{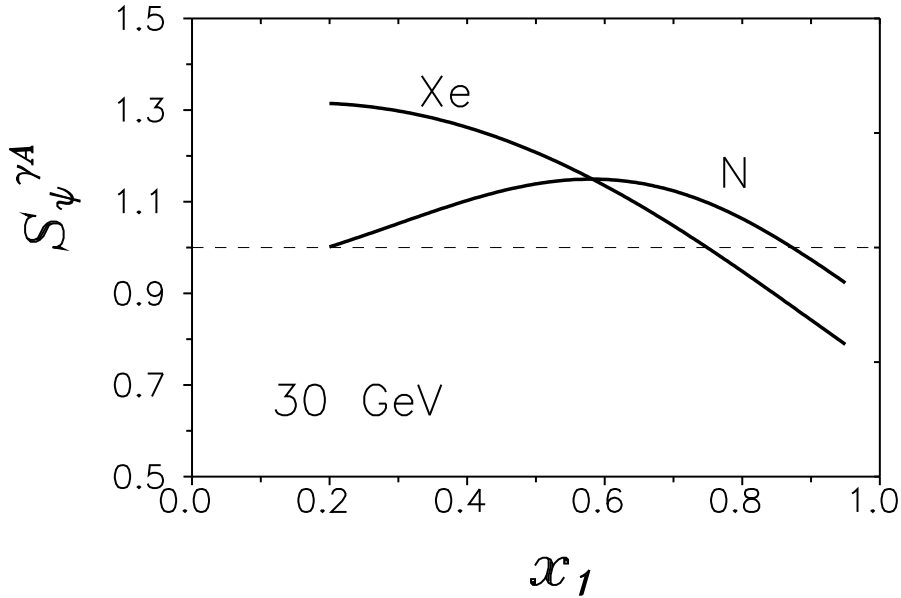}
\begin{center}
\vspace{7.5cm}
\parbox{13cm}
 {\caption[Delta]
 {Nuclear suppression/enhancement factor for $J/\Psi$ 
photoproduction on xenon and nitrogen at $\nu=30\ GeV$}
\label{fig5}}
\end{center}
\end{figure}

\noi 
{\large\bf 7. Conclusions and discussion}
\medskip

We summarize the main results of the paper as follows.
\begin{itemize}

\item
The cross section of inelastic photoproduction of charmonia
off nuclei is calculated within the eikonal (Glauber)
approximation. Corresponding formulas are derived 
for the first time. 
We also correct the Glauber formulas for inelastic shadowing,
which makes nuclear medium substantially more transparent
for charmonium.

\item
 The experimentally observed
 nuclear enhancement of
 inelastic
$J/\Psi$ photoproduction at $x_1 < 0.85$
is found to be too large compared with expectation of 
Glauber model to be explained by an 
enhancement of gluons in nuclei at low
 $x_{Bj}$.  
 
\item
A novel QCD 
 mechanism of
 charmonium photoproduction is suggested,
 which is not included in the
 standard
 multiple-scattering theory and can only 
occur in a nuclear target.  This
 mechanism arises as a natural
 consequence of the colored structure of
 the
 Pomeron: the projectile hadron
 (or a hadronic fluctuation of the photon)
 induces a color dipole
 in the target nucleus, leaving the
 nucleus in a
 colorless state.
 This is a
 specific way of diffractive excitation
 of the
 nucleus and inclusive production
 of leading hadrons.
 
\item
As different from the Glauber model the new mechanism
contribution is essentially energy-dependent
and violates Feynman scaling. Due to steeper an A-dependence
this mechanism successfully competes with Glauber one
at moderate $x_1$ in the energy range of the EMC and NMC
experiments and nicely explains the observed
\cite{emc,nmc} nuclear
 enhancement of the
photoproduction cross
 section.
This effect is expected to be squeezed towards 
$x_1=1$ and to vanish under the quasielastic peak
at higher energies.

\item
 We conclude that the observed nuclear
 enhancement of inelastic
photoproduction
 of $J/\Psi$ is the evidence of a new
 phenomenon related to
specifics of QCD.
 An analogous effect in the case of
 interaction of light
hadrons turns out
 to be suppressed and escapes observations
 \cite{kl,lnpi}.
Photoproduction of
 heavy quarkonia might be a unique
 reaction where the new
mechanism can be
 observed.
 
\end{itemize}

\noi
 Note, that a specific channel of decay
 of the color-dipole created inside
the
 nucleus into two nucleons was suggested
 in \cite{kn} as a source of the
backward
 protons in the laboratory frame.
 Measurements by the SIGMA
Collaboration
 \cite{sigma} of backward proton
 production in pion-beryllium
interaction
 at $40\ GeV$ nicely confirmed the
 theoretical expectations.
This is
 another manifestation of
 the of double-color-exchange mechanism.

 The important signature of the mechanism
 under discussion its specific
energy
 dependence.  The nuclear enhancement
 does not vanish at higher
energies, but
 is squeezed towards $x_1 = 1$. This is
 different from gluon
distribution in a
 nucleus, which anyway have to switch to
 nuclear shadowing
at higher energies,
 i.e.  smaller Bjorken $x_2$.
\medskip

\noi
 {\bf
 Acknowledgements:} We are thankful to A.~Br\"ull and Ch.~Mariotti
providing us with detailed results of the NMC experiment for stimulating
discussion and J.~Nemchik for useful comments. This work was partially
 supported by a grant from the Federal Ministry for Education and
Research (BMBF), grant number 06~HD~742, and by INTAS grant 93-0239.
\bigskip

\noi
 {\large\bf Appendix}
\medskip

We estimate the correction $\delta(d\sigma^{\gamma
A}_{\Psi}/dx_1)$ from inelastic production of $J/\Psi$
with inelasticity $\alpha$ at one point in the nucleus,
with subsequent inelastic rescattering with inelasticity
$\beta$ on another bound nucleon. This contribution to the
cross section integrated over transverse momenta
reads

 \beq
 \delta\left( \frac{d\sigma^{\gamma
A}_{\Psi}}{dx_1}\right) \approx
A\ \widetilde T\ 
 \int\limits_0^1
d\alpha\ \int\limits_0^1
 d\beta\ \delta(x_1-\alpha\beta)\
 \frac{d\sigma^{\gamma N}_{\Psi}}{d\alpha}\
\frac{d\sigma^{\gamma N}_{\Psi}}{d\beta}\ ,
\label{19}
\eeq
 
 \noi
 where
 
 \beq
 \widetilde T =
{1\over 2A} \int d^2b\
 T^2(b)\exp[-\sigma_{in}^{\Psi N} T(b)]
\label{20}
\eeq
 
We assume for the sake of simplicity that the energy is
high, $q_{el}R_A \ll 1$.  

 After integration (\ref{19}) takes
 the form
 
 \beq
\delta\left( \frac{d\sigma^{\gamma
 A}_{\Psi}}{dx_1}\right) =
\widetilde T\ K(x_1)\ A\ 
 \frac{d\sigma^{\gamma
N}_{\Psi}}{dx_1}\ ,
\label{21}
\eeq
 
 noi
 where the factor $K(x_1)$ is a slow
 function of $x_1$,
 
\beq
 K(x_1) =
\left(\frac{\sigma^{\Psi p}_{tot}}
 {\sigma^{pp}_{tot}}\right)^2\
\frac{G(pp \to pX)}{B^{\Psi p}_{in}}\ 
\left[(1+x_1)\
 \ln\left(\frac{1-x_1}{\epsilon}\right)
 +x_1\ ln\left({1\over
x_1}
 \right)\right]
\label{22}
\eeq

\noi
Here $\epsilon = M^2_0/s$, where $M_0$ is the minimal mass
of the excited nucleon in final state, which we include
in the integration over inelasticities in (\ref{19}).
At $M_0 = 2\ GeV$ and $x_1 = 0.9$ we estimate the factor 
$K \approx 0.03\ fm^2$ 

The effective nuclear thickness (\ref{20}) for tin is 
$\widetilde T^{Sn} \approx 0.3\ fm^{-2}$.

We conclude that according to these estimates and expression
(\ref{22}) the correction under discussion has nearly the same
form of $x_1$-dependence as the standard mechanism contribution,
but is suppressed by the factor $K\ \widetilde T\ A/S^{\gamma N}_{\Psi}
\approx 0.016$. We can neglect such a small 
correction for our purposes.

\setlength{\baselineskip} {20pt}

\end{document}